\documentstyle [12pt,eqsecnum,aps] {revtex}
\input epsf
\newcommand{\beq}{\begin{equation}}
\newcommand{\eeq}{\end{equation}}
\newcommand{\beqs}{\begin{eqnarray}}
\newcommand{\eeqs}{\end{eqnarray}}

\begin{document}

\draft

\baselineskip 6.0mm

\title{Lower Bounds and Series for the Ground State Entropy of the 
Potts Antiferromagnet on Archimedean Lattices and their Duals}

\vspace{4mm}

\author{Robert Shrock\thanks{email: shrock@insti.physics.sunysb.edu}
\and Shan-Ho Tsai\thanks{email: tsai@insti.physics.sunysb.edu}}

\address{
Institute for Theoretical Physics  \\
State University of New York       \\
Stony Brook, N. Y. 11794-3840}

\maketitle

\vspace{4mm}

\begin{abstract}

   We prove a general rigorous lower bound for 
$W(\Lambda,q)=\exp(S_0(\Lambda,q)/k_B)$, the exponent of the ground state 
entropy of the $q$-state Potts antiferromagnet, on an arbitrary Archimedean 
lattice $\Lambda$.  We calculate large-$q$ series expansions for the exact 
$W_r(\Lambda,q)=q^{-1}W(\Lambda,q)$ and compare these with our lower bounds 
on this function on the various Archimedean lattices.  It is shown that the 
lower bounds coincide with a number of terms in the large-$q$ expansions and 
hence serve not just as bounds but also as very good approximations to the 
respective exact functions $W_r(\Lambda,q)$ for large $q$ on the various 
lattices $\Lambda$.  Plots of $W_r(\Lambda,q)$ are given, and the general 
dependence on lattice coordination number is noted.  Lower bounds and series 
are also presented for the duals of Archimedean lattices.  As 
part of the study, the chromatic number is determined for all Archimedean 
lattices and their duals. Finally, we report calculations of chromatic zeros 
for several lattices; these provide further support for our earlier 
conjecture that a sufficient condition for $W_r(\Lambda,q)$ to be analytic at 
$1/q=0$ is that $\Lambda$ is a regular lattice. 

\end{abstract}

\pacs{05.20.-y, 64.60.C, 75.10.H}

\vspace{10mm}

\pagestyle{empty}
\newpage

\pagestyle{plain}
\pagenumbering{arabic}
\renewcommand{\thefootnote}{\arabic{footnote}}
\setcounter{footnote}{0}

\section{Introduction}

    Nonzero ground state disorder and associated entropy, $S_0 \ne 0$, is an
important subject in statistical mechanics.  One physical example is provided
by ice, for which the residual molar entropy is 
$S_0 = 0.82 \pm 0.05$ cal/(K-mole), i.e., $S_0/R = 0.41 \pm 0.03$, where 
$R=N_{Avog.}k_B$ \cite{ice}-\cite{atkins}.  Indeed, residual entropy at low
temperatures has been observed in a number of substances, including nitrous 
oxide, NO, carbon monoxide, CO, and FClO$_3$ (a comprehensive review is given 
in Ref. \cite{ps}).  In these 
examples, the entropy occurs without frustration, i.e., the configurational 
energy can be minimized.  In magnetic systems, two examples are provided by the
Ising antiferromagnet on the triangular and kagom\'e lattices
\cite{wannier,kn}; here, the ground state entropy does involve frustration.
A particularly simple model exhibiting 
ground state entropy without the complication of frustration is the $q$-state 
Potts antiferromagnet (AF) \cite{potts} on a lattice $\Lambda$, for 
$q \ge \chi(\Lambda)$, where $\chi(\Lambda)$ denotes the minimum number of
colors necessary to color the vertices of the lattice such that no two adjacent
vertices have the same color.  As is already evident from the foregoing, this 
model also has a deep connection with 
graph theory in mathematics \cite{fk}-\cite{rtrev}, since the 
zero-temperature partition function of 
the above-mentioned $q$-state Potts antiferromagnet on a lattice $\Lambda$ 
satisfies $Z(\Lambda,q,T=0)_{PAF}=P(\Lambda,q)$, where $P(G,q)$ is the 
chromatic 
polynomial expressing the number of ways of coloring the vertices of a graph 
$G$ with $q$ colors such that no two adjacent vertices (connected by a bond of
the graph) have the same color; hence, the ground state entropy per site is
given by $S_0/k_B = \ln W(\Lambda,q)$, where $W(\Lambda,q)$, the ground state
degeneracy per site, is 
\beq 
W(\Lambda,q) = \lim_{n \to \infty} P(\Lambda_n,q)^{1/n}
\label{w}
\eeq
Here, $\Lambda_n$ denotes an $n$-vertex lattice of type $\Lambda$ (square,
triangular, etc.), with 
appropriate (e.g., free) boundary conditions. Given the above connection, 
it is convenient to express our bounds on the ground state entropy in terms 
of its exponent, $W(\Lambda,q)$.  Since nontrivial exact solutions for this
function are known in only a very few cases (square lattice for $q=3$
\cite{lieb}, triangular lattice \cite{baxter}, and kagom\'e lattice for $q=3$
\cite{wurev}), it is important to exploit and extend general approximate 
methods that can be applied to all cases.  
Such methods include rigorous upper and lower 
bounds, large-$q$ series expansions, and Monte Carlo measurements.  Recently, 
we studied the ground state entropy in antiferromagnetic Potts models on 
various lattices and obtained further results with these three methods 
\cite{p3afhc}-\cite{wa}. 

   In the present paper we achieve a substantial generalization of our previous
studies. Among other things, we obtain a general rigorous lower bound on the 
(exponent of the) ground state entropy that applies for all 
Archimedean lattices (for definitions, see below) and find further examples of 
lattices where this lower bound coincides to many orders with a 
large-$q$ series expansion of the respective $W$ function.  This agreement is
particularly striking since, {\it a priori}, a lower bound on a function need
not coincide with any, let alone many, of the terms in the series expansion of
the function about a given point.  We also present calculations of chromatic
zeros for a number of lattices and show that they support a conjecture that we
have made earlier \cite{wa}.  The reader is referred to Refs. 
\cite{p3afhc}-\cite{wa} for further background and references. 

   Although the full set of Archimedean lattices is not as much discussed in 
the physics literature as the subset of
three homopolygonal (i.e., monohedral) ones, viz., square, triangular, and 
honeycomb, other Archimedean lattices are of both theoretical and
experimental interest.  For example, the kagom\'e lattice is of current
interest because of the experimental observation of compounds whose behavior
can be modelled by quantum Heisenberg antiferromagnets on this lattice, 
including SrCr$_{8-x}$Ga$_{4+x}$O$_{19}$, (SCGO(x)) \cite{scgo} and, recently, 
deuteronium jarosite, (D$_3$O)Fe$_3$(SO$_4)_2$(OD)$_6$ \cite{dj}.  Related to
this, one of the reasons for interest in the kagom\'e lattice is that on this
lattice the quantum Heisenberg antiferromagnet has a disordered ground state
with finite entropy and frustration \cite{hafkag}.  Indeed, it has been known
for a number of years that although the Ising antiferromagnet exhibits ground 
state entropy, frustration, and zero long range order on both the triangular 
and kagom\'e lattices, 
the greater degree of disorder on the kagom\'e lattice is evidenced 
by the fact that while the correlation length diverges as $T \to 0$ on the 
triangular lattice \cite{steph}, it remains finite on the kagom\'e lattice 
\cite{suto}. 

\section{Some Graph Theory Background}

   In this section we include some basic definitions and results in graph
theory that will be necessary for our work.  A graph (with no loops or
multiple bonds) is defined as a collection of vertices and edges (bonds) 
joining various vertices.  In strict mathematical terminology, a graph 
involves a finite number of vertices; the regular (infinite) lattices which we
will consider here are thus limits of graphs, with some appropriate (e.g. free)
boundary conditions.  The chromatic polynomial $P(G,q)$, defined above, was 
first introduced by Birkhoff \cite{birk} and has been the subject of 
intensive mathematical study since \cite{whit}-\cite{bl}, 
\cite{harary}-\cite{rtrev}. 
The zeros of the chromatic polynomial $P(G,q)$ are denoted the chromatic 
zeros of $G$.  A graph $G$ that can be colored with $q$ colors, i.e., has 
$P(G,q) > 0$, is said to 
be $q$-colorable.  The chromatic number $\chi(G)$ of a graph $G$ is the 
minimum (integer) $q$ such that one can color the graph subject to the 
condition that no two adjacent vertices have the same color, i.e. such that 
$P(G,q) > 0$. The graph $G$ is then said to be $\chi(G)$-chromatic.  
The graph is uniquely $\chi(G)$-chromatic if and 
only if $P(G,q=\chi(G))=q!$. A uniquely $q$-chromatic 
graph is $q$-partite.  These definitions can be
extended to the regular infinite lattices of interest here if one defines the
latter as the limit as the number of vertices $n \to \infty$ of finite-$n$
lattice graphs with appropriate boundary conditions, such as free boundary
conditions.  For bipartite lattices one can also use
periodic boundary conditions that preserve the bipartite property.  
Similarly, for a tripartite lattice, one can use periodic boundary 
conditions that preserve the tripartite property, etc.   Henceforth, when we
refer to a finite lattice of type $\Lambda$, it is understood that appropriate
boundary conditions are specified.  

\begin{figure}
\centering
\leavevmode
\epsfxsize=4.0in
\epsffile{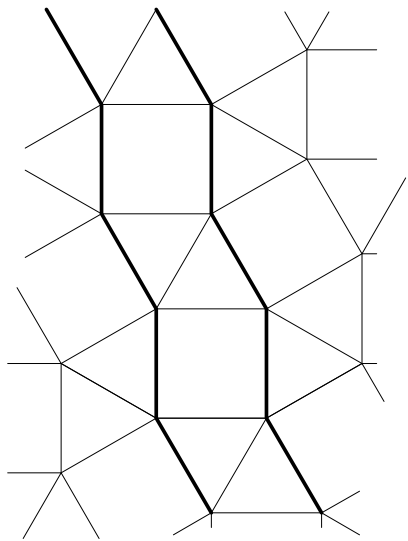}
\caption{Section of the $(3^2 \cdot 4 \cdot 3 \cdot 4)$ Archimedean
lattice with paths ${\cal L}_n$ and ${\cal L}_n'$ used in the proof of our
lower bound indicated with darker lines. See text for discussion.}
\label{fig33434}
\end{figure}

\begin{figure}
\centering
\leavevmode
\epsfxsize=4.0in
\epsffile{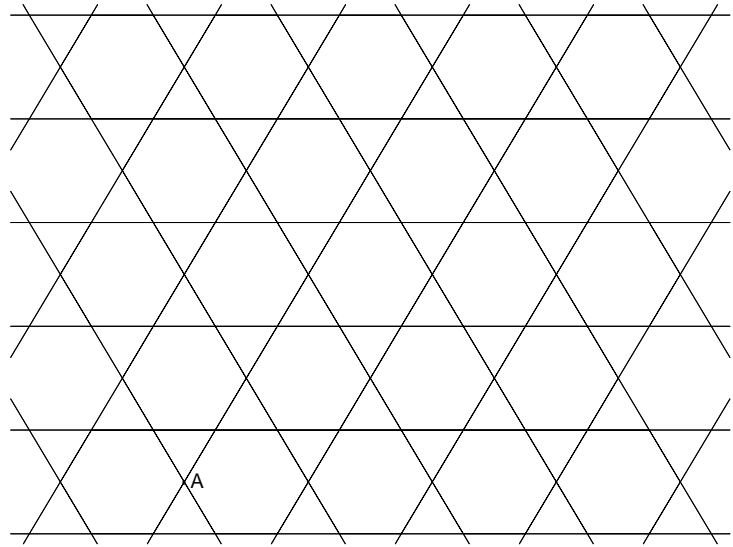}
\caption{Section of the $(3 \cdot 6 \cdot 3 \cdot 6)$ Archimedean lattice. The 
point A labels a vertex that is referred to later in the text.}
\label{fig3636}
\end{figure}

\begin{figure}
\centering
\leavevmode
\epsfxsize=4.0in
\epsffile{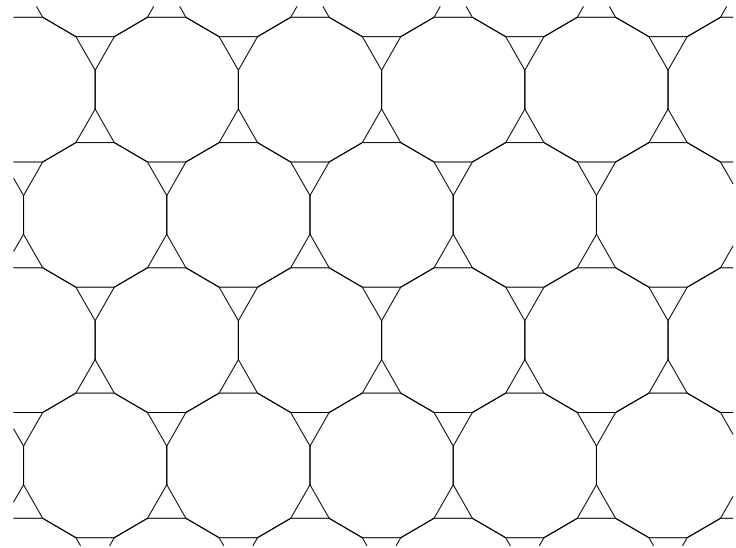}
\caption{Section of the $(3 \cdot 12^2)$ Archimedean lattice.} 
\label{fig31212}
\end{figure}

\begin{figure}
\centering
\leavevmode
\epsfxsize=4.0in
\epsffile{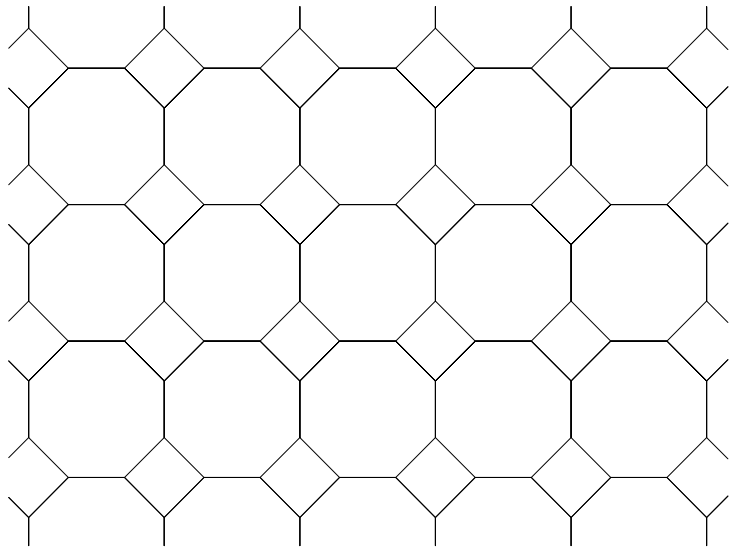}
\caption{Section of the $(4 \cdot 8^2)$ Archimedean lattice.} 
\label{fig488}
\end{figure}

  An Archimedean lattice is defined as a uniform tiling of the plane by 
regular polygons in which all vertices are equivalent \cite{gsbook}.  Such a
lattice is specified by the ordered sequence of polygons that one traverses in
making a complete circuit around a vertex in a given (say counterclockwise)
direction.  This is incorporated in the mathematical notation for an 
Archimedean lattice $\Lambda$: 
\beq
\Lambda = (\prod_i p_i^{a_i}) 
\label{lambda}
\eeq
where in the above circuit, the notation $p_i^{a_i}$ indicates that the 
regular polygon $p_i$ occurs contiguously $a_i$ times; it can also occur
noncontiguously.  Because the starting point is irrelevant, the symbol is 
invariant under cyclic permutations. For later purposes, when a
polygon $p_i$ occurs several times in a non-contiguous manner in the product,
we shall denote $a_{i,s}$ as the sum of the $a_i$'s over all of the
occurrences of the given $p_i$ in the product.   There are eleven Archimedean
lattices:
\beqs
\{\Lambda \} & = & \{(3^6), \ (4^4), \ (6^3), \ (3^4 \cdot 6), \ 
(3^3 \cdot 4^2), \ (3^2 \cdot 4 \cdot 3 \cdot 4), \ 
(3 \cdot 4 \cdot 6 \cdot 4), \\ \cr
& & (3 \cdot 6 \cdot 3 \cdot 6), \ 
(3 \cdot 12^2), \ (4 \cdot 6 \cdot 12), \ (4 \cdot 8^2) \}
\label{archset}
\eeqs
Of these lattices, three are homopolygonal, also called monohedral, i.e., 
they only involve one
type of regular polygon: $(3^6)$ (triangular), $(4^4)$ (square), and $(6^3)$ 
(hexagonal or honeycomb).  The other eight are heteropolygonal, i.e., involve 
tilings with more than one type of regular polygon.  The kagom\'e lattice is 
$(3 \cdot 6 \cdot 3 \cdot 6)$. Some illustrative pictures
of the heteropolygonal lattices are helpful for understanding the results of
the present paper.  Accordingly, we show in Figs. \ref{fig33434}-\ref{fig488}
the $(3^2 \cdot 4 \cdot 3 \cdot 4)$, $(3 \cdot 6 \cdot 3 \cdot 6)$ (kagom\'e),
$(3 \cdot 12^2)$, and $(4 \cdot 8^2)$ lattices.  Pictures of others are given 
in Ref. \cite{gsbook}; see also Ref. \cite{cmo}.  

The geometric constraint
that the internal angles of each polygon adjacent to a given vertex sum to
$2\pi$ is 
\beq
\sum_i a_{i,s}\Bigl (1 - \frac{2}{p_i} \Bigr ) = 2
\label{heterocon}
\eeq
The solutions to this equation for integral $a_{i,s}$ and $p_i$, together with
the requirement that a local patch be extendable to tile the plane, yield the
Archimedean lattices. 
The degree $\Delta$ of a vertex of a graph $G$ is the number of edges (bonds)
that connect to this vertex.  For a regular (infinite) lattice, this is the 
same as the coordination number.  For an Archimedean lattice (\ref{lambda}),
the coordination number is 
\beq
\Delta = \sum_i a_{i,s}
\label{delta}
\eeq
Of course, for a finite lattice with free boundary conditions, the vertices 
on the 
boundary have lower values of $\Delta$ than those in the interior; this will 
not be important for our rigorous bounds, which pertain to the thermodynamic
limit on an infinite lattice.  For a homopolygonal lattice 
$\Lambda=(p^a)$, the constraint (\ref{heterocon}) relates the coordination 
number to $p$ according to 
\beq
\Delta = a = \frac{2p}{p-2} \ , \qquad {\rm for} \quad \Lambda = (p^a)
\label{delta_homo}
\eeq
which can also be written in the symmetric form $\Delta^{-1} + p^{-1} = 1/2$. 
The girth $g(G)$ of a graph $G$ is the length of a minimum circuit on the 
graph. Hence, for an Archimedean lattice (\ref{lambda}), $g = \min\{p_j \}$. 
The number of polygons of type $p_i$ per site is given by
\beq
\nu_{p_i} = \frac{N_{p_i \ per \ v}}{N_{v \ per \ p_i}} = \frac{a_{i,s}}{p_i}
\label{nu}
\eeq
The set of homopolygonal Archimedean lattices is invariant under the duality
transformation, which interchanges 0-cells (vertices) and 2-cells (faces) and
thus maps $(p^a) \to (a^p)$.  When one applies the duality transformation
to the other eight Archimedean lattices, the resultant lattices are not
Archimedean.

\begin{figure}
\centering
\leavevmode
\epsfxsize=4.0in
\epsffile{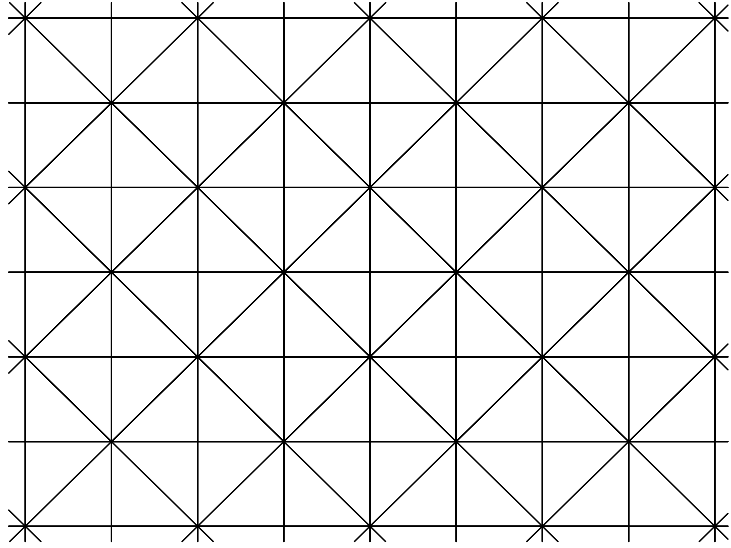}
\caption{Section of the $[4 \cdot 8^2]$ dual Archimedean (Laves) lattice.}
\label{fig488lav}
\end{figure}

     The dual of the Archimedean lattice, often called a Laves lattice
\cite{gsbook,laves}, is defined by listing the ordered sequence of vertex 
types specified by their degrees, $v_i = \Delta_i$, along the boundary 
of any polygon:
\beq
\Lambda_{Laves} = \Bigl [ \prod_i v_i^{b_i} \Bigr ]
\label{laves}
\eeq
where in the above product, the notation $v_i^{b_i}$ indicates that the
vertex $v_i$ with degree $v_i=\Delta_i$ occurs contiguously $b_i$ times; it 
can also occur noncontiguously.  As with Archimedean lattices, because the 
starting point is irrelevant, the symbol is invariant under cyclic 
permutations. When a vertex of type $v_i$ occurs several times in a 
non-contiguous manner in the product, we shall denote $b_{i,s}$ as the sum of 
the $b_i$'s over all of the occurrences of the given $v_i$ in the product.   
There are eleven dual Archimedean lattices. 
Just as all vertices are equivalent on an Archimedean lattice, all faces are
equivalent on its dual; these are comprised of a single type of polygon, $p$, 
which, however, does not in general have sides of equal length.  Note that 
\beq
p = \sum_i b_{i,s}
\label{plav}
\eeq
and the girth $g([\prod_i v_i^{b_i}]) = p$. The dual of the Archimedean 
lattice $(\prod_i p_i^{a_i})$ is $[\prod_i v_i^{b_i}]$ with $v_i=p_i$ and 
$b_i = a_i$.  Among Laves lattices, three involve only vertices of one type and
are denoted homovertitial and are equivalent to the respective three
homopolygonal Archimedean lattices; these are $[3^6]=(6^3)$ (honeycomb),
$[4^4]=(4^4)$ (square), and $[6^3]=(3^6)$ (triangular); the other eight are
heterovertitial, i.e. involve vertices of at least two types.  An example of a
heterovertitial Laves lattice is $[4 \cdot 8^2]$, the union-jack lattice, 
shown in Fig. \ref{fig488lav}.  

For these duals of Archimedean lattices, the formula for the number of 
$p$-gons per site is 
\beq
\nu_p = \biggl [ \sum_i \frac{b_{i,s}}{v_i} \biggr ]^{-1}
\label{nuplav}
\eeq
Although different vertices on a heterovertitial Laves lattice have different
degrees, it will be useful for later purposes to introduce the notion of an 
effective degree or coordination number, defined by 
$\Delta_{eff} = \lim_{V \to \infty} \frac{2E}{V}$, where $V$ and $E$ denote 
the vertices and edges of the lattice graph.  This is given by 
\beq
\Delta_{eff} = \nu_p p 
\label{deltaefflav}
\eeq
This completes our brief review of necessary
definitions from graph theory. 

\section{Chromatic Numbers for Archimedean and Dual Archimedean Lattices}

   As part of our work on ground state entropy, it has been useful to 
calculate the chromatic numbers $\chi(\Lambda)$ for the Archimedean lattices 
and their duals, and to determine whether or not these lattices are uniquely 
$q$-chromatic for $q=\chi(\Lambda)$. As background, 
we recall that from the four-color theorem \cite{ah}, by duality, it follows
that $\chi(G) \le 4$ for any planar graph \cite{ah}. We find that for all of
the nonbipartite Archimedean and dual Archimedean lattices, $\chi=3$ except for
the $[3 \cdot 12^2]$ Laves lattice, for which $\chi=4$.  The latter result
follows from the fact that this lattice contains $K_4$ subgraphs, and
$\chi(K_m)= m$.  (Here, $K_m$ is called the complete graph on $m$ vertices,
defined such that each vertex is joined by bonds to every other vertex.)  We
summarize our results \cite{lit} in Tables \ref{archtable} and
\ref{lavestable}, together with some properties of these
lattices that will be relevant to our calculations of bounds and series. 

\eject

\begin{table}
\caption{Chromatic number $\chi(\Lambda)$ and determination of uniqueness or
non-uniqueness of coloring if $q=\chi(\Lambda)$ for the 
Archimedean lattices.  UQC stands for ``uniquely $q$-chromatic'', Y,N for yes
and no.}
\begin{center}
\begin{tabular}{ccc}
$\Lambda$ & $\chi(\Lambda)$ & UQC \\
$(3^6)$                         & 3 & Y \\
$(4^4)$                         & 2 & Y \\
$(6^3)$                         & 2 & Y \\ 
$(3^4 \cdot 6)$                 & 3 & Y \\
$(3^3 \cdot 4^2)$               & 3 & N \\
$(3^2 \cdot 4 \cdot 3 \cdot 4)$ & 3 & N \\
$(3 \cdot 4 \cdot 6 \cdot 4)$   & 3 & N \\
$(3 \cdot 6 \cdot 3 \cdot 6)$   & 3 & N \\
$(3 \cdot 12^2)$                & 3 & N \\
$(4 \cdot 6 \cdot 12)$          & 2 & Y \\
$(4 \cdot 8^2)$                 & 2 & Y \\
\hline
\end{tabular}
\end{center}
\label{archtable}
\end{table}

\begin{table}
\caption{Chromatic number $\chi(\Lambda)$ and determination of uniqueness or
non-uniqueness of coloring if $q=\chi(\Lambda)$ for the heterovertitial 
duals of Archimedean lattices.  UQC stands for ``uniquely
$q$-chromatic''.} 
\begin{center}
\begin{tabular}{ccc}
$\Lambda$                       & $\chi(\Lambda)$     & UQC \\
$[6^3]$                         & 3                   & Y \\
$[4^4]$                         & 2                   & Y \\
$[3^6]$                         & 2                   & Y \\ 
$[3^4 \cdot 6]$                 & 3                   & N \\
$[3^3 \cdot 4^2]$               & 3                   & N \\
$[3^2 \cdot 4 \cdot 3 \cdot 4]$ & 3                   & N \\
$[3 \cdot 4 \cdot 6 \cdot 4]$   & 2                   & Y \\
$[3 \cdot 6 \cdot 3 \cdot 6]$   & 2                   & Y \\
$[3 \cdot 12^2]$                & 4                   & N \\
$[4 \cdot 6 \cdot 12]$          & 3                   & Y \\
$[4 \cdot 8^2]$                 & 3                   & Y \\
\hline
\end{tabular}
\end{center}
\label{lavestable}
\end{table}

\section{A Rigorous Lower Bound on $W(\Lambda,\lowercase{q})$ for Archimedean
$\Lambda$ } 

   In Refs. \cite{ww,w3}, we derived rigorous lower and upper bounds on 
$W(\Lambda,q)$ for the triangular and honeycomb lattices, 
using a coloring matrix method of the type first applied by Biggs to obtain 
such bounds for the square lattice \cite{biggs77}.  
We showed that these upper and lower bounds rapidly 
approached each other for large $q$ and became very restrictive even for
moderately large $q$.  Furthermore, we found that for a wide range of $q$
values, the lower bounds were very close to the respective actual values of 
$W(\Lambda,q)$. Accordingly, here we shall focus on rigorous lower bounds for 
$W(\Lambda,q)$ for the general class of Archimedean lattices, of which the
homopolygonal lattices are a special case.  We shall derive a general
lower bound applicable to any Archimedean lattice.  In addition, we shall
derive lower bounds for the duals of Archimedean lattices. 

   Before proceeding, it is necessary to recall a subtlety in the definition of
the function $W(\Lambda,q)$.  As we discussed in Ref. \cite{w}, the formal 
eq. (\ref{w}) is not, in general, adequate to define $W(\Lambda,q)$ because 
of a noncommutativity of limits
\beq
\lim_{n \to \infty} \lim_{q \to q_s} P(\Lambda,q)^{1/n} \ne
\lim_{q \to q_s} \lim_{n \to \infty} P(\Lambda,q)^{1/n}
\label{wnoncomm}
\eeq
at certain special points $q_s$.  At these points, one must also specify the
order of the limits in (\ref{wnoncomm}). We denote these definitions as
\beq
W(\{G\},q_s)_{D_{qn}} \equiv \lim_{q \to q_s} \lim_{n \to \infty} P(G,q)^{1/n}
\label{wdefqn}
\eeq
and 
\beq
W(\{G\},q_s)_{D_{nq}} \equiv \lim_{n \to \infty} \lim_{q \to q_s} P(G,q)^{1/n}
\label{wdefnq}
\eeq
where $\{G\}$ denotes the $n \to \infty$ limit of the family of $n$-vertex
graphs of type $G$.
One can maintain the analyticity of $W(\{G\},q)$ at the special points $q_s$ 
of $P(G,q)$ by choosing the order of limits in (\ref{wdefqn}); however, 
this definition produces a function $W(\{G\},q)_{D_{qn}}$ whose values at the 
points $q_s$ differ significantly from the values which one would get for 
$P(G,q_s)^{1/n}$ with finite-$n$ graphs $G$.  The definition based on the 
opposite order of limits, (\ref{wdefnq}) 
gives the expected results like $W(\{G\},q_s)=0$ for $q_s=0,1$,
and, for $G \supseteq \triangle$ (i.e., for $G$ containing at least one 
triangle as a subgraph), also $q_s=2$.  However, this second definition 
yields a function $W(\{G\},q)$ with discontinuities at the set of points
$\{q_s\}$.  Following Ref. \cite{w}, in our results below, in order to avoid 
having to write special formulas for the points $q_s$, we shall adopt the 
definition $D_{qn}$ but at appropriate places will take note of the 
noncommutativity of limits (\ref{wnoncomm}).  As will be evident from the
derivation, our rigorous lower bounds are on the function 
$W(\Lambda,q) \equiv W(\Lambda,q)_{D_{qn}}$, and they apply for 
the range of (positive integer) $q$ values for which the relevant coloring 
matrices (see below) are nontrivial, i.e., for $q \ge \chi(\Lambda)$; the 
values of $\chi(\Lambda)$ were listed in Tables \ref{archtable} and 
\ref{lavestable}. 

   Clearly, a general upper bound on a chromatic polynomial for an $n$-vertex
graph $G$ is $P(G,q) \le q^n$.  This yields the
corresponding upper bound $W(\{G\},q) < q$.  Hence, it is natural to define a
reduced function that has a finite limit as $q \to \infty$, 
\beq
W_r(\{G\},q) = q^{-1}W(\{G\},q)
\label{wr}
\eeq
When calculating large-$q$ Taylor series expansions for $W$ functions on
regular lattices , it is most convenient to carry this out for the related
function 
\beq
\overline W(\Lambda,y) = \frac{W(\Lambda,q)}{q(1-q^{-1})^{\Delta/2}} 
\label{wbar}
\eeq
for which the large-$q$ series can be written in the form 
\beq
\overline W(\Lambda,y)=1+\sum_{m=1}^\infty w_{\Lambda,m} y^m
\label{wseries}
\eeq
with 
\beq
y = \frac{1}{q-1}
\label{y}
\eeq
For duals of Archimedean lattices one can use the same formulas with the
replacement $\Delta \to \Delta_{eff}$, where $\Delta_{eff}$ was given in
eq. (\ref{deltaefflav}). 

   Our rigorous lower bounds are of the form 
\beq
W(\Lambda,q) \ge W(\Lambda,q)_\ell 
\label{wlb}
\eeq
where the subscript $\ell$ denotes ``lower'', and we shall give the explicit
expressions for $W(\Lambda,q)_\ell$ below.  We shall use two other equivalent
forms of the bounds, namely on the functions $W_r(\Lambda,q)$ and $\overline
W(\Lambda,y)$, both of which have the finite limit $W_r(\Lambda,q=\infty) =
\overline W(\Lambda,y=0)=1$ and hence are convenient to compare with large-$q$
(small-$y$) series.  Thus, for the latter function, the bounds read 
\beq
\overline W(\Lambda,y) \ge \overline W(\Lambda,y)_\ell \ 
\label{wbarlb}
\eeq
where the reduced lower bound function is defined by analogy with (\ref{wbar})
as 
\beq
\overline W(\Lambda,y)_\ell = \frac{W(\Lambda,q)_\ell}{q(1-q^{-1})^{\Delta/2}} 
\label{wbarylb}
\eeq

   Our general rigorous lower bound, which is a major result of the present
paper, is proved in the following theorem:

\vspace{3mm}

\begin{flushleft}

Theorem

\end{flushleft}

Let $\Lambda=(\prod_i p_i^{a_i})$ be an Archimedean lattice. Then for (integer)
$q \ge \chi(\Lambda)$, $W(\Lambda,q) \equiv W(\Lambda,q)_{D_{qn}}$ has the 
lower bound
\beq
W \Bigl ( (\prod_i p_i^{a_i}),q \Bigr )_\ell = 
\frac{\prod_i D_{p_i}(q)^{\nu_{p_i}}}{q-1}
\label{wlbarch}
\eeq
where the $\{i\}$ in the product label the set of $p_i$-gons involved
in $\Lambda$, $\nu_{p_i}$ was defined in eq. (\ref{nu}), and 
\beq
D_k(q) = \frac{P(C_k,q)}{q(q-1)} = 
\sum_{s=0}^{k-2}(-1)^s {{k-1}\choose {s}} q^{k-2-s}
\label{dk}
\eeq
where $P(C_k,q)=(q-1)^k+(-1)^k(q-1)$ is the chromatic polynomial for a
$k$-vertex circuit graph, i.e., polygon. 
This lower bound takes a somewhat simpler form in 
terms of the related function $\overline W(\Lambda,y)_\ell$:
\beq
\overline W \Bigl ( (\prod_i p_i^{a_i}),y \Bigr )_\ell = 
\prod_i \Bigl [ 1+(-1)^{p_i}y^{p_i-1} \Bigl ]^{\nu_{p_i}}
\label{wbarlbarch}
\eeq

\vspace{4mm}

\begin{flushleft}

Proof

\end{flushleft} 

    We consider a sequence of finite 2D Archimedean lattices with 
periodic boundary conditions in one direction, say the $x$ direction, 
and either periodic or free boundary conditions in the orthogonal, $y$, 
direction.  Denote the lengths of the finite lattice in these two directions as
$m$ and $n$ and the finite lattice of type $\Lambda$ as $\Lambda_{m\times n}$.
Extending the method that Biggs used for the square lattice \cite{biggs77}, we 
introduce a coloring matrix $T$, somewhat analogous to the transfer matrix 
for statistical mechanical spin models.  The construction of $T$ begins by 
considering an $n$-vertex path ${\cal L}_n$ along a vertical path on
$\Lambda_{m \times n}$.  For the square lattice, it is obvious what is meant 
here, and one can also represent the other homopolygonal lattices 
as square lattices with bonds added or deleted 
(for the triangular and honeycomb lattices,
respectively; see Fig. 1 of Ref. \cite{ww} and the discussion in
Ref. \cite{w3}).  Thus, as is well known, the triangular lattice can be 
deformed so that it is represented as a square lattice with bonds added to 
connect the lower left and upper right vertices of each square, and the 
honeycomb lattice can be deformed to make a brick lattice, with the long side
of the bricks oriented vertically. These deformations 
do not affect the coloring properties of the lattices.  With these
representations of the triangular and honeycomb lattices, one can choose a
vertical path in the same manner as for the square lattice. Analogous paths 
can be defined on heteropolygonal Archimedean lattices (an illustration is
given in Fig. \ref{fig33434} for the $(3^2 \cdot 4 \cdot 3 \cdot 4)$ lattice).
In all cases except the
$(4 \cdot 6 \cdot 12)$ lattice, these are simple (i.e., unbranched) chains.  
For the $(4 \cdot 6 \cdot 12)$ lattice, the paths are chains with single-bond 
branches.  We shall give the proof first for the seven Archimedean lattices 
where the paths are simple chains which do not intersect, namely, the 
three homopolygonal lattices and the $(3^2 \cdot 4^2)$, 
$(3^2 \cdot 4 \cdot 3 \cdot 4)$, $(3 \cdot 4 \cdot 6 \cdot
4)$, and $(4 \cdot 8^2)$ heteropolygonal lattices.  We then give the proof for
the three remaining lattices where the paths do intersect, and, separately, for
the special case of the $(4 \cdot 6 \cdot 12)$, where the paths, although
non-intersecting, are branched chains.  The number of allowed
colorings of the path ${\cal L}_n$ is $P({\cal L}_n,q)= q(q-1)^{n-1} \equiv 
{\cal N}$. Now focus on two adjacent paths ${\cal L}_n$ and ${\cal L}_n'$.  
Define compatible $q$-colorings of these paths as colorings such that no two 
vertices $v \in {\cal L}_n$ and $v' \in {\cal L}_n'$ connected by a bond of 
$\Lambda_{m \times n}$ have the same color.  One can then associate with this 
pair of paths an ${\cal N} \times {\cal N}$ dimensional symmetric matrix $T$ 
with entries $T_{{\cal L}_n,{\cal L}_n'}=1$ or 0 if the $q$-colorings of 
${\cal L}_n$ and ${\cal L}_n'$ are or are not compatible, respectively.  
Then for fixed lengths of the lattice in the $x$
and $y$ directions, $m$ and $n$, $P(\Lambda_{m \times n},q)=Tr(T^m)$.  For a
given $n$, since $T$ is a nonnegative matrix, one can apply the
Perron-Frobenius theorem \cite{pf} to conclude that $T$ has a real positive
eigenvalue $\lambda_{max,n}(q)$.  Hence, for fixed $n$, 
\beq
\lim_{m \to \infty} Tr(T^m)^{1/(mn)} \to \lambda_{max}^{1/n}
\label{trlim}
\eeq
so that
\beq
W(\Lambda,q) = \lim_{n \to \infty} \lambda_{max}^{1/n}
\label{wlim}
\eeq
Denote the column sum $\kappa_j(T) = \sum_{i=1}^{\cal N} T_{ij}$ (equal to
the row sum $\rho_j(T)=\sum_{i=1}^{\cal N}T_{ji}$ since $T^T=T$) and
$S(T) = \sum_{i,j=1}^{\cal N} T_{ij}$; note that $S(T)/{\cal N}$ is the 
average row (column) sum.  The lower bound is then a consequence of the 
($r=1$ special
case of the) theorem that for a nonnegative symmetric matrix $T$ \cite{london} 
$\lambda_{max}(T) \ge [S(T^r)/{\cal N}]^{1/r}$ for $r=1, 2,...$ together with
eq. (\ref{wlim}): 
\beq
W(\Lambda,q) \ge W(\Lambda,q)_\ell = 
\lim_{n \to \infty} \biggl (\frac{S(T)}{\cal N} \biggr )^{1/n}
\label{lowbound}
\eeq
This is the general method; we next proceed to calculate $S(T)$. To do this we
observe that the adjacent paths ${\cal L}_n$ and ${\cal L}_n'$ can be 
chosen such that the
strip-like section of the lattice between them contains each of the types of
polygons comprising $\Lambda$ (this essentially amounts to the orientation of
the lattice to define the vertical direction).  Each polygon $p_i$ has the
chromatic polynomial given above as $P(C_{p_i},q)$.  Next, we use a basic 
theorem from graph theory: if $G$ and $G'$ are graphs that intersect in a 
complete graph $K_r$ \ \cite{comp}, then 
$P(G \cup G',q)=P(G,q)P(G',q)/P(K_r,q)$, where
$P(K_r,q)=\prod_{s=0}^{r-1}(q-s)$.  We then apply this theorem, taking into 
account that the number of polygons of type $p_i$ per vertex is $\nu_{p_i}$, 
to obtain the result that for paths of length $n$, $S(T)
= c(q)\prod_i D_{p_i}(q)^{\nu_{p_i}(n + b)}$, where $b$ is an unimportant 
constant independent of $n$, and the prefactor $c(q)=q(q-1)(q-2)$ if 
$\Lambda$ contains triangles, $\Lambda \supseteq \bigtriangleup$, and 
$c(q)=q(q-1)$ otherwise.  
Then taking the $n \to \infty$ limit in (\ref{lowbound}) yields
eq. (\ref{wlbarch}).  Equivalently, using the definition of the reduced
function (\ref{wbar}), we obtain (\ref{wbarlbarch}).  For the three 
lattices, viz., $(3^4 \cdot 6)$, $(3 \cdot 6 \cdot 3 \cdot 6)$, and 
$(3 \cdot 12^2)$, where the adjacent paths 
${\cal L}_n$ and ${\cal L}_n'$ do intersect, the calculation involves a 
technical complication due to these intersections: the expression for $S(T)$
involves an additional factor of $(q-1)^{\lambda n}$, where $\lambda$ is the
fraction of points on each path ${\cal L}_n$ that coincide with points on
${\cal L}_n'$, but this factor is exactly cancelled by the same additional
factor in the expression for ${\cal N}$ in the denominator of the ratio
$S(T)/{\cal N}$.  The additional factor in the denominator arises because one
must correct for the undercounting of the points on the paths ${\cal L}_n$ due
to the intersections.  Because the factor cancels, the result is the same for
as for the lattices with nonintersecting adjacent paths ${\cal L}_n$ and ${\cal
L}_n'$.  

   Finally, we consider the $(4 \cdot 6 \cdot 12)$ lattice.  Let us consider
the lattice as oriented so that the 12-gons have vertical left and right bonds
and horizontal top and bottom bonds, with a hexagon vertically above the 12-gon
and adjacent to it, and label the vertices of each 12-gon in a clockwise 
manner starting with the left vertex on the top bond.  The path
${\cal L}_n$ that we use goes from vertex 6 of a given 12-gon to vertices 
5, 4, 3, and then crosses over to vertex 8 of the adjacent 12-gon in
the upper right (``northeast'') direction, has a 1-bond branch to vertex 7 of
this 12-gon, and then continues along on vertices 8 through 12 and then 1, with
a 1-bond branch to vertex 2, continuing from vertex 1 upward to vertex 6 of the
vertically adjacent 12-gon in the northwest direction, and so forth.  
The strip enclosed by two adjacent paths of length $n$ of this type has a
chromatic polynomial 
\beq
S(T) = q(q-1)\Bigl [ D_4(q)^3D_6(q)^2D_{12}(q) + 2D_3(q)D_4(q)^2D_6(q) 
+ qD_3(q)^2D_4(q) \Bigr ]^{n/12}
\label{s4612}
\eeq
The chromatic polynomial for the path of length $n$ is the same as that for an
unbranched path, since both are tree graphs.  Hence, for this case, the lower 
bound actually reads $W((4 \cdot 6 \cdot 12),q) \ge 
W((4 \cdot 6 \cdot 12),q)_{\ell'}$, where 
\beq
W((4 \cdot 6 \cdot 12),q)_{\ell'} = \frac{\Bigl [ D_4(q)^3D_6(q)^2D_{12}(q)
+ 2D_3(q)D_4(q)^2D_6(q) + qD_3(q)^2D_4(q) \Bigr ]^{1/12}}{q-1}
\label{fulllb4612}
\eeq
For the above-mentioned range of $q$ under consideration here, i.e., $q \ge
\chi((4 \cdot 6 \cdot 12))=2$, the two additional terms in the
square brackets are positive, so that 
\beq
W((4 \cdot 6 \cdot 12),q)_\ell = 
\frac{D_4(q)^{1/4}D_6(q)^{1/6}D_{12}(q)^{1/12}}{q-1} 
\label{lb4612}
\eeq
which is of the form (\ref{wlbarch}).  (As we shall discuss below, the
difference between the bounds (\ref{fulllb4612}) and (\ref{lb4612}) is very
small.)  This completes the proof. 

\vspace{4mm} 

For the homopolygonal Archimedean lattices $\Lambda = (p^a)$ (with $a=\Delta$
given by (\ref{delta}) and $\nu_p=2/(p-2)$ from (\ref{nu})), our bound 
(\ref{wlbarch}) reduces to 
\beq
W \Bigl ( (p^\Delta),q \Bigr )_\ell = \frac{D_{p}(q)^{2/(p-2)}}{q-1}
\label{wlb_homo}
\eeq
or equivalently, (\ref{wbarlbarch}) reduces to 
\beq
\overline W \Bigl ( (p^\Delta),y \Bigr )_\ell = 
\Bigl [ 1 + (-1)^p y^{p-1} \Bigr ]^{2/(p-2)}
\label{wbarlb_homo}
\eeq
In Table \ref{wbartable}, we list the explicit forms of the lower bounds 
$\overline W(\Lambda,y)_\ell$ for the Archimedean lattices, including a 
comparison with small-$y$ series, to be discussed below. 

\begin{table}
\caption{Rigorous lower bounds $\overline W(\Lambda,y)_\ell$ for Archimedean 
lattices $\Lambda=(\prod_i p_i^{a_i})$ given by eq. (\ref{wbarlbarch}). 
For homopolygonal Archimedean lattices
$\Lambda=(p^\Delta)$, the Taylor series expansion of 
$\overline W(\Lambda,y)_\ell$ coincides to order $O(y^{i_c})$ with the 
series expansion of $\overline W(\Lambda,y)$, where $i_c=i_{max}=2(p-1)$, the
values of which are listed in the third column, and the subscript $c$ stands
for ``coinciding''.  For other lattices, the
series expansions of $\overline W(\Lambda,y)_\ell$ and 
$\overline W(\Lambda,y)$ coincide to at least order $O(y^{i_c})$.}
\begin{center}
\begin{tabular}{ccc}
$\Lambda$ & $\overline W(\Lambda,y)_\ell$ & $i_c$ \\
$(3^6)$   & $(1-y^2)^2$       & 4  \\
$(4^4)$   & $1+y^3$           & 6  \\
$(6^3)$   & $(1+y^5)^{1/2}$   & 10 \\ 
$(3^4 \cdot 6)$ & $(1-y^2)^{4/3}(1+y^5)^{1/6}$ & 4  \\ 
$(3^3 \cdot 4^2)$               & $(1-y^2)(1+y^3)^{1/2}$  & 4  \\
$(3^2 \cdot 4 \cdot 3 \cdot 4)$ & $(1-y^2)(1+y^3)^{1/2}$  & 4 \\
$(3 \cdot 6 \cdot 3 \cdot 6)$  & $(1-y^2)^{2/3}(1+y^5)^{1/3}$ & 8 \\
$(3 \cdot 4 \cdot 6 \cdot 4)$   & $(1-y^2)^{1/3}(1+y^3)^{1/2}(1+y^5)^{1/6}$ 
& 5 \\
$(3 \cdot 12^2)$   &  $(1-y^2)^{1/3}(1+y^{11})^{1/6}$  & 13 \\ 
$(4 \cdot 6 \cdot 12)$     & $(1+y^3)^{1/4}(1+y^5)^{1/6}(1+y^{11})^{1/12}$ 
& 11 \\
$(4 \cdot 8^2)$            & $(1+y^3)^{1/4}(1+y^7)^{1/4}$ & 12 \\ 
\hline
\end{tabular}
\end{center}
\label{wbartable}
\end{table}

   We give some illustrations of the method of the proof here.  First, we
illustrate this for a lattice where the paths ${\cal L}_n$ do not intersect, 
namely, the $(3^2 \cdot 4 \cdot 3 \cdot 4)$ lattice shown in
Fig. \ref{fig33434}.  The paths ${\cal L}_n$ and ${\cal L}_n'$ are depicted 
by the thicker lines, and the strip between these consists of a sequence of 
squares and double triangles.  For technical convenience, we consider a path to
start at the lower lefthand vertex of a square, and we take $n$ to be odd, so
that the path length is even. 
The sum $S(T)$ is calculated starting from the basic graph $G_{433}$ 
comprised of a square and two adjacent triangles (say above the square) for 
which the chromatic polynomial is $P(G_{433},q)=q(q-1)D_3(q)^2D_4(q)$.  For 
a strip lying between the adjacent paths ${\cal L}_n$ and ${\cal L}_n'$ (both
of length $n-1$), starting the count from the lower left corner of a
square, there are $r=(n-1)/2$ $G_{433}$ graphs, so, in an obvious notation, 
\beq
S(T) = P(G_{433}^r,q) = q(q-1)[D_3(q)^2D_4(q)]^{(n-1)/2}
\label{s433m}
\eeq
Dividing by ${\cal N}=q(q-1)^{n-1}$ and taking the $n \to \infty$ limit of the
$1/n$'th power of the ratio as in eq. (\ref{lowbound}), one obtains the
rigorous lower bound
\beq
W((3^2 \cdot 4 \cdot 3 \cdot 4),q)_\ell = \frac{D_3(q)D_4(q)^{1/2}}{q-1}
\label{lb33434}
\eeq
which is seen to be the special case of (\ref{wlbarch}) for this lattice. 
Another explicit example is provided by the $(4 \cdot 8^2)$ lattice, which we 
have previously discussed \cite{w3}.  

   To illustrate the proof in a case where the paths ${\cal L}_n$ and 
${\cal L}_n'$ have a nonzero overlap, we discuss the $(3 \cdot 6 \cdot 3 \cdot
6)$ lattice, shown in Fig. \ref{fig3636}.  To visualize the path ${\cal L}_n$,
start at the vertex A at which two triangles touch and move 
``northwest'' to the
leftmost vertex of the hexagon above these triangles; then move northeast to
the vertex common to the two triangles above the hexagon, then northwest again
to the leftmost vertex of the next higher hexagon, etc., continuing upward in
this zigzag manner.  For ${\cal L}_n'$, start from the same vertex, but move
first northeast to the rightmost vertex of the hexagon above the triangles,
then northwest to the vertex common to the two triangles above the hexagon, and
so forth.  For technical convenience, we consider the paths to be of length a 
multiple of 4. 
Evidently, the $n$-vertex paths ${\cal L}_n$ and ${\cal L}_n'$ so
defined have a nonzero overlap (intersection) consisting of the vertices that
are shared in common by the pairs of triangles traversed along the route. For
consistency, we assign alternate intersection points to be on adjacent paths. 
These intersection points comprise 1/3 of the total number $n$ of the 
vertices along each path.  The
strip between the paths ${\cal L}_n$ and ${\cal L}_n'$ consists of a sequence
of triangle-hexagon-triangle graphs, which we denote as $G_{363}$.  For a path
${\cal L}_n$ starting from a vertex of type A, there are $r=(n-1)/3$ linked 
$G_{363}$ graphs in this strip. The chromatic polynomial of the strip is 
\beq
S(T) = P(G_{363}^r,q) = q[(q-1)D_3(q)^2D_6(q)]^{(n-1)/3}
\label{s363}
\eeq
However, in evaluating the denominator for eq. (\ref{lowbound}), it is
necessary to correct for the vertices that are common to both paths; taking
into account that these comprise a third of the total points on each path and 
including this correction factor yields the denominator 
${\cal N} = q(q-1)^{4n/3}$. Evaluating eq. (\ref{lowbound}) then gives the
resulting lower bound
\beq
W((3 \cdot 6 \cdot 3 \cdot 6)_\ell = \frac{D_3(q)^{2/3}D_6(q)^{1/3}}{q-1}
\label{wlb3636}
\eeq
which again is the special case of (\ref{wlbarch}) for this lattice. 
  
   Finally, we discuss the difference between the lower bounds 
(\ref{fulllb4612}) and (\ref{lb4612}) for the $(4 \cdot 6 \cdot 12)$
lattice. This difference shows that, at least for this lattice, the special
case (\ref{lb4612}) is not an optimal lower bound.  However, the numerical
difference is extremely small.  At $q=\chi((4 \cdot 6 \cdot 12))=2$, the bounds
(\ref{fulllb4612}) and (\ref{lb4612}) are identical and equal to 1, which is
also the exact value of $W((4 \cdot 6 \cdot 12),q=2)_{D_{nq}}$ (see eq. 
(\ref{wbip1}) below).  At $q=3$, the difference between the bounds 
(\ref{fulllb4612}) and (\ref{lb4612}) is $1.3 \times 10^{-5}$, and this
difference decreases monotonically for larger $q$.  This reflects the fact that
the large-$q$ series expansions of the corresponding reduced lower bound
functions are identical to $O(q^{-14})$: 
\beq
W_r((4 \cdot 6 \cdot 12),q)_{\ell'} - W_r((4 \cdot 6 \cdot 12),q)_{\ell} = 
\frac{1}{6}q^{-15} + O(q^{-16}) 
\label{wdif4612}
\eeq

   We comment on some general properties of our rigorous lower bound 
on $W(\Lambda,q)$ in eq. (\ref{wlbarch}), and the consequent bound on the
related functions $W_r(\Lambda,q)$ and $\overline W(\Lambda,y)$ for 
Archimedean lattices $\Lambda$. First, 
\beq
W \Bigl ( (\prod_i p_i^{a_i}),q \Bigr )_\ell \ \sim \ 
\frac{q^{\sum_i \nu_i(p_i-2)}}{q} \ \sim \ q \ , \qquad {\rm as} \quad 
q \to \infty
\label{wlbasymp}
\eeq
where $\sum_i \nu_i(p_i-2)=2$ as a consequence of eq. (\ref{heterocon}).  
Hence, $\lim_{q \to \infty} W_r\Bigl ( (\prod_i p_i^{a_i}),q \Bigr )_\ell = 1$.

   Second, note that when we compare our lower bound on 
$W_r(\Lambda,q)$ to the large-$q$ Taylor 
series expansion on each lattice $\Lambda$, this comparison is obviously 
restricted to the range of $q$ for which these series expansions are 
applicable.  As we discussed in Ref. \cite{w},
for a given graph $G$, when one generalizes $q$ to a complex variable, the
reduced function $W_r(\{G\},q)=q^{-1}W(\{G\},q)$ is analytic in the $q$ 
plane except on
the union of boundaries ${\cal B}$.  These boundaries can separate the $q$
plane into various regions $R_i$ such that one cannot analytically continue
$W(\{G\},q)$ from one region to another. We defined the region containing the 
positive real $q$ axis from a minimal value $q_c(\Lambda)$ to $q=\infty$ as
$R_1$.  Clearly, the large-$q$ Taylor series for $W_r(\{G\},q)$ apply in this 
region.  Here, $q_c(G)$ is thus defined as the maximal (finite) real
value of $q$ where $W(\{G\},q)$ is nonanalytic. All of these considerations 
apply
for the case of regular lattice graphs $\Lambda$ defined, as above, as the 
$n \to \infty$ limit of finite $\Lambda_n$ lattices.  As our general study and
explicit exact solutions in Ref. \cite{w} showed for $q \in R_1$,
the limits (\ref{wnoncomm}) commute and hence the two definitions 
(\ref{wdefqn}) and (\ref{wdefnq}) coincide.  Hence, in our discussions of
series, we shall not need to distinguish between these two different
definitions of $W(\Lambda,q)$.  Furthermore, in this region $R_1$, just as one
can extend the definition of the function $W(\Lambda,q)$ from integer to real
$q$, so also one can carry out the same extension of the lower bound
$W(\Lambda,q)_\ell$.  Hence, for $q \ge q_c(\Lambda)$, the lower bound
(\ref{wlbarch}) applies for real, and not just integer $q$. 
In general,
\beq
\chi(\Lambda) \le q_c(\Lambda)
\label{qcchi}
\eeq
In Refs. \cite{p3afhc} and \cite{w}, we showed that 
$q_c((3^6))=4$, $q_c((4^4))=3$, and $q_c((6^3))=(3+\sqrt{5})/2$. 

   Third, $W(\Lambda,q)_\ell$ is a monotonically increasing function 
of real $q$ for $q \ge \chi(\Lambda)$ \ \cite{wred}.

   Fourth, if two different lattices involve the same set of polygons and 
have the property that each type of polygon $p_i$ occurs an equal total 
number of times as one makes the circuit around each vertex, then our lower 
bound (\ref{wlbarch}) is the same for both.  This is true of the $(3^3 \cdot
4^2)$ and $(3^2 \cdot 4 \cdot 3 \cdot 4)$ lattices, for which the lower bound
is given by (\ref{lb33434}).

   Fifth, it should be remarked that one can often choose different types of
paths ${\cal L}_n$ and ${\cal L}_n'$. For example, in Ref. \cite{ww} we derived
a lower bound $W((6^3),q)_{\ell, dimer} = (q-1)^{3/2}/q^{1/2}$ by choosing
paths consisting of certain dimers.  In Ref. \cite{w3}, using paths of the type
described here, we obtained the lower bound of the form (\ref{wlbarch}), 
\beq
W((6^3),q)_\ell = \frac{D_6(q)^{1/2}}{q-1} \ge W((6^3),q)_{\ell,dimer}
\label{whc}
\eeq
which, as indicated, is slightly more stringent.  

  Finally, the fact that for the circuit with an even number, say $2k$, of 
vertices, the chromatic
polynomial is $P(C_{2k},q=2)=2$, together with the definition (\ref{dk}), it
follows that \cite{dk0} 
\beq
D_{2k}(q=2)=1
\label{d2k}
\eeq
Hence, for bipartite Archimedean lattices $\Lambda_{bip.}$, where our lower 
bound (\ref{wlbarch}) can be applied at $q=\chi(\Lambda_{bip.})=2$, we have 
\beq
W(\Lambda_{bip.},q=2)_\ell=1
\label{wlbbip1}
\eeq
Thus, in this case, although our bound refers to  $W(\Lambda,q)_{D_{qn}}$, it
has the same value as $W(\Lambda_{bip.},q=2)_{D_{nq}}$, as given below in
eq. (\ref{wbip1}).

   Note that for lattices $\Lambda_{UQC}$ that are uniquely $q$-chromatic
(UQC) for $q=\chi(\Lambda)$, it follows, {\it a fortiori}, that 
\beq
W(\Lambda_{UQC},q=\chi(\Lambda))_{D_{nq}} = 1
\label{wuqc1}
\eeq
Thus, in particular, for a bipartite lattice
\beq
W(\Lambda_{bip.},q=2)_{D_{nq}}=1
\label{wbip1}
\eeq
and for the tripartite lattices $(3^6)=[6^3]$, $(3^4 \cdot 6)$, $[4 \cdot
8^2]$, and $[4 \cdot 6 \cdot 12]$, 
\beq
W({\Lambda_{trip.}},q=3)_{D_{nq}}=1.
\label{wtrip1}
\eeq

\section{A Rigorous Lower Bound on $W(\Lambda,\lowercase{q})$ for Dual 
Archimedean $\Lambda$ }

   Using methods similar to those that we used for the Archimedean lattices, 
we have also calculated rigorous lower bounds for the duals of these lattices. 
As with the lower bounds for Archimedean lattices, our bounds for the dual
lattices again apply for $q \ge \chi(\Lambda)$ for each $\Lambda$.
For seven of the lattices, we are able to obtain the respective lower
bounds by using paths ${\cal L}_n$ that either do not intersect, in the case of
the homovertitial lattices and also for $[3^4 \cdot 6]$, $[3^3 \cdot 4^2]$, 
$[3^2 \cdot 4 \cdot 3 \cdot 4]$, and $[3 \cdot 4 \cdot 6 \cdot 4]$, or
intersect, in the case of the $[3 \cdot 6 \cdot 3 \cdot 6]$ lattice.  For
these, by the same methods as before, we obtain the general rigorous lower
bound
\beq
W([\prod_i v_i^{b_i}],q)_\ell = \frac{D_p(q)^{\nu_p}}{q-1}
\label{wlblav}
\eeq
where $p$ and $\nu_p$ were given in eqs. (\ref{plav}) and 
(\ref{nuplav}). 
Using eq. (\ref{wbarylb}) with $\Delta$ replaced by $\Delta_{eff}$ given in 
eq. (\ref{deltaefflav}), this is equivalent to 
\beq
\overline W([\prod_i v_i^{b_i}],q)_\ell = 
\Bigl ( 1+(-1)^p y^{p-1}\Bigr )^{\nu_p}
\label{wbarlblav}
\eeq
The proof is the same as the one given before for Archimedean lattices, with
the simplification that only one kind of polygon is involved in the calculation
of $S(T)$. 
The homovertitial Laves lattices have already been dealt with as homopolygonal
Archimedean lattices; for the heterovertitial Laves lattices satisfying the
above condition on paths we find the specific bounds 
\beq
W([3^4 \cdot 6],q)_\ell = W([3^3 \cdot 4^2],q)_\ell = 
W([3^2 \cdot 4 \cdot 3 \cdot 4],q)_\ell = \frac{D_5(q)^{2/3}}{q-1}
\label{w33344lavlb}
\eeq
and 
\beq
W([3 \cdot 6 \cdot 3 \cdot 6],q)_\ell = 
W([3 \cdot 4 \cdot 6 \cdot 4],q)_\ell = \frac{D_4(q)}{q-1} = 
\frac{(q^2-3q+3)}{q-1}
\label{w3636lavlb}
\eeq
\beq
W([4 \cdot 8^2],q)_{\ell,{\cal L}} = \frac{D_3(q)^2}{q-1} = \frac{(q-2)^2}{q-1}
\label{w488lavlb1}
\eeq
We note that the lower bound (\ref{w3636lavlb}) is the same as the one that we
derived for the square lattice $(4^4)=[4^4]$, and the lower bound
(\ref{w488lavlb1}) is the same as the one that we derived for the triangular
lattice, $(3^6)=[6^3]$.

However, for one of the lattices, viz., $[4 \cdot 8^2]$, where such paths 
exist, we have been able to obtain a more stringent lower bound by using a 
different kind of coloring matrix method, in which this matrix is defined in 
terms of the compatibility not of adjacent paths, but of adjacent chains of 
graphs that are more complicated than paths.  Consider the $[4 \cdot 8^2]$ 
lattice shown in Fig. \ref{fig488lav}.  
For this lattice, one obtains the lower 
bound (\ref{w488lavlb1}). But rather than paths, one can consider
chains of graphs constructed as follows: consider a vertical line on the
lattice, and let ${\cal M}_n$ denote the chain of triangles whose bases include
vertices from 1 to $n$ along this vertical line and whose apexes point to the
right.  Evidently, each pair of sequential triangles intersect at the common
vertex on their 
bases.  Define ${\cal M}_n'$ to be the corresponding equal-length chain of 
triangles whose bases form the unit segments along the next (adjacent) 
vertical line of the lattice to the immediate right of the previous one. 
For technical convenience, let $n$ be odd. 
Then define a new type of (symmetric) coloring matrix, of dimension 
${\cal N} \times {\cal N}$, where ${\cal N} = P({\cal M}_n,q)= 
P({\cal M}_n',q)$, 
with entries $T_{{\cal M}_n,{\cal M}_n'}=1$ or 0 if the $q$-colorings of the
chains of graphs ${\cal M}_n$ and ${\cal M}_n'$ are or are not compatible, 
respectively.  One then proceeds as before, to derive the lower bound
(\ref{lowbound}).  In the present case,
\beq
P({\cal M}_n,q) = q[(q-1)(q-2)]^r \ , \quad r = \frac{n-1}{2}
\label{ptriangles488}
\eeq
and 
\beq
S(T) = q(q-1)[(q-2)^2(q^2-5q+7)]^r \ , \quad r = \frac{n-1}{2}
\label{st}
\eeq
so that, evaluating the limit in (\ref{lowbound}), we obtain
\beq
W([4 \cdot 8^2],q)_\ell = \biggl [ \frac{(q-2)(q^2-5q+7)}{(q-1)} \biggr ]^{1/2}
\label{w488lavlb}
\eeq
For the relevant range, $q \ge \chi([4 \cdot 8^2])=3$, the lower bound
(\ref{w488lavlb}) always lies above (\ref{w488lavlb1}), i.e., is more
stringent.  We list the corresponding lower bound 
$\overline W([4 \cdot 8^2],y)_\ell$ in Table \ref{wbardualtable}). 

   Using this different coloring matrix method based on chains of graphs that
are not simple paths, we similarly obtain the following rigorous lower bounds
for the other two dual Archimedean lattices: 
\beq
W([3 \cdot 12^2],q)_\ell = \frac{\Bigl [(q-2)(q-3) \Bigr ]^{2/3} }{(q-1)^{1/3}}
\label{w31212lavlb}
\eeq
\beq
W([4 \cdot 6 \cdot 12],q)_\ell = \frac{(q-2)(q^2-5q+7)^{1/3}}{(q-1)^{2/3}}
\label{w4612lb}
\eeq

The corresponding lower bounds $\overline W(\Lambda,y)_\ell$ are listed in
Table \ref{wbardualtable}.  To retain the symmetry between the Archimedean and
Laves lattices, we also list the lower bounds for the square, triangular, and 
honeycomb lattice in their respective Laves forms, $[4^4]$, $[6^3]$, and 
$[3^6]$.

\begin{table}
\caption{Rigorous lower bounds $\overline W(\Lambda,y)_\ell$ for duals of 
Archimedean lattices, $\Lambda=[\prod_i v_i^{b_i}]$, where $v_i = \Delta_i$. 
For homovertitial duals, $\Lambda=[v^b]=[\Delta^p]$ where $p$ and $\Delta$ are
related by eq. (\ref{delta_homo}), the Taylor series expansion of 
$\overline W(\Lambda,y)_\ell$ coincides to order $O(y^{i_c})$ with the 
series expansion of $\overline W(\Lambda,y)$, where $i_c=i_{max}=2(p-1)$, the
values of which are listed in the third column, and the subscript $c$ stands
for ``coinciding''.  For other lattices, the
series expansions of $\overline W(\Lambda,y)_\ell$ and 
$\overline W(\Lambda,y)$ coincide to at least order $O(y^{i_c})$.}
\begin{center}
\begin{tabular}{ccc}
$\Lambda$ & $\overline W(\Lambda,y)_\ell$ & $i_c$ \\
$[3^6]$   & $(1+y^5)^{1/2}$       & 10  \\
$[4^4]$   & $1+y^3$               &  6  \\
$[6^3]$   & $(1-y^2)^2$           &  4 \\ 
$[3^4 \cdot 6]$                  & $(1-y^4)^{2/3}$ & 7 \\
$[3^3 \cdot 4^2]$                & $(1-y^4)^{2/3}$ & 7 \\ 
$[3^2 \cdot 4 \cdot 3 \cdot 4]$  & $(1-y^4)^{2/3}$ & 7 \\
$[3 \cdot 6 \cdot 3 \cdot 6]$    & $1+y^3$         & 4 \\ 
$[3 \cdot 4 \cdot 6 \cdot 4]$    & $1+y^3$         & 4 \\
$[3 \cdot 12^2]$             & $(1+y)^2(1-3y+2y^2)^{2/3}$ & 3 \\ 
$[4 \cdot 6 \cdot 12]$   & $(1+y)^2(1-y)(1-3y+3y^2)^{1/3}$ & 3 \\ 
$[4 \cdot 8^2]$          & $(1+y)^2(1-y)^{1/2}(1-3y+3y^2)^{1/2}$ & 3 \\ 
\hline
\end{tabular}
\end{center}
\label{wbardualtable}
\end{table}

\section{Large-$\lowercase{q}$ Series}

   {\it A priori}, a lower bound on a given function need not agree to any, let
alone many, terms in the Taylor series expansion of that function about some
special point.  In our earlier comparison of rigorous lower bounds on
$\overline W(\Lambda,y)$ with large-$q$ (i.e., small-$y$) series expansions 
for the respective exact $\overline W(\Lambda,y)$ functions for the 
homopolygonal lattices $\Lambda=(p^\Delta)$, where $\Delta = 2p/(p-2)$, we 
found \cite{w,ww,w3} that our lower bounds and the series coincided to order 
$O(y^{i_c})$ where 
\beq
i_c = i_{max} = 2(p-1)
\label{imaxhomo}
\eeq
In the case of the $(6^3)$ (honeycomb) lattice, our lower bound thus coincided
with the small-$y$ series for the exact function to $O(y^{10})$, i.e., the
first eleven terms. Thus, in addition to being a rigorous lower bound, 
our expression for $\overline W((6^3),y)$ became an extremely accurate
approximation to the exact function, $\overline W((6^3),y)$ for even moderately
large $q$.  We also considered one heteropolygonal Archimedean lattice, the 
$(4 \cdot 8^2)$ lattice, and calculated both a lower bound and small-$y$ series
for $\overline W((4 \cdot 8^2),y)$ \cite{w3}; in this case we found that our 
lower bound coincided with the first thirteen terms of our series.  

   Here we report our calculations of small-$y$ series for $\overline
W(\Lambda,y)$ on the remaining seven Archimedean lattices and compare these 
with our rigorous lower bounds, thereby achieving a complete comparison for all
Archimedean lattices.  Our calculations use the methods of Ref. \cite{nagle}. 
Since our main purpose was an exploratory study of the extent to which our
lower bounds coincided with the series expansions, we have not attempted to
compute the latter to very high order; it is straightforward to carry these
expansions to higher order (e.g. \cite{lat312}). 

Our results are listed below: 
\beq
\overline W((3^4 \cdot 6),y) = 1 - \frac{4}{3}y^2 + \frac{2}{9}y^4 + O(y^5)
\label{wbar33336taylor}
\eeq
\beq
\overline W((3^3 \cdot 4^2),y) =
\overline W((3^2 \cdot 4 \cdot 3 \cdot 4),y) = 1 - y^2 + \frac{y^3}{2} 
+ 0 \cdot y^4 + O(y^5) 
\label{wbar33434taylor}
\eeq
\beq
\overline W((3 \cdot 4 \cdot 6 \cdot 4),y) = 1 - \frac{y^2}{3}
+ \frac{y^3}{2} - \frac{y^4}{9} + 0 \cdot y^5 + O(y^6) 
\label{wbar3464taylor}
\eeq
\beq
\overline W((3 \cdot 6 \cdot 3 \cdot 6),y) = 1 - \frac{2}{3}y^2
- \frac{1}{9}y^4 + \frac{1}{3}y^5 - \frac{4}{3^4}y^6 - \frac{2}{9}y^7
-\frac{7}{3^5}y^8 + O(y^9)
\label{wbar3636taylor}
\eeq
\beq
\overline W((3 \cdot 12^2),y) = 1 - \frac{1}{3}y^2 - \frac{1}{9}y^4
- \frac{5}{3^4}y^6 - \frac{10}{3^5}y^8 - \frac{22}{3^6}y^{10} +
\frac{1}{6}y^{11} - \frac{154}{3^8}y^{12} - \frac{1}{18}y^{13} + O(y^{14})
\label{wbar31212taylor}
\eeq
\beq
\overline W((4 \cdot 6 \cdot 12),y) = 1 + \frac{1}{4}y^3 +
\frac{1}{6}y^5 - \frac{3}{32}y^6 + \frac{1}{24}y^8 + \frac{7}{128}y^9
- \frac{5}{72}y^{10} + \frac{13}{192}y^{11} + O(y^{12})
\label{wbar4612taylor}
\eeq
\beq
\overline W((4 \cdot 8^2),y) = 1 + \frac{1}{4}y^3 - \frac{3}{2^5}y^6
+ \frac{1}{4}y^7 + \frac{7}{2^7}y^9 + \frac{1}{2^4}y^{10}
- \frac{77}{2^{11}}y^{12} + O(y^{13})
\label{wbar488taylor}
\eeq

   We have also calculated low-order series for dual Archimedean lattices in 
order to make an initial comparison with our rigorous lower bounds. 
We have compared each of these series with the corresponding small-$y$
Taylor series expansions of our rigorous lower bounds in Tables \ref{wbartable}
and \ref{wbardualtable}.  We find that the latter
expansions coincide at least to the order $O(y^{i_c})$, where the respective 
values of $i_c$ are listed in Table \ref{wbartable} for each lattice. 
As is evident from this table, the
striking fact that we showed previously \cite{ww,w3}, viz., that the series
expansions for $\overline W(\Lambda,y)_\ell$ coincide to very high order with
the respective series expansions of the exact functions 
$\overline W(\Lambda,y)$ for $\Lambda=(6^3)$ (to $O(y^{10})$) and for 
$\Lambda=(4 \cdot 8^2)$ (to at least $O(y^{12})$), is not restricted to just 
these 
lattices.  Indeed, we observe that for both of the lattices $(3 \cdot 12^2)$
and $(4 \cdot 6 \cdot 12)$, the respective small-$y$ series for the lower bound
and the exact function coincide to at least $O(y^{13})$ and $O(y^{11})$,
respectively \cite{lat312}.  Thus, in general, this
establishes that for a number of lattices our rigorous lower bounds actually 
serve as quite good approximations to the exact $W$ functions, especially for
large $q$. 

\section{Plots of $W_{\lowercase{r}}(\Lambda,\lowercase{q})$ and 
$\overline W(\Lambda,\lowercase{y})$ }

\begin{figure}
\centering
\leavevmode
\epsfxsize=4.0in
\epsffile{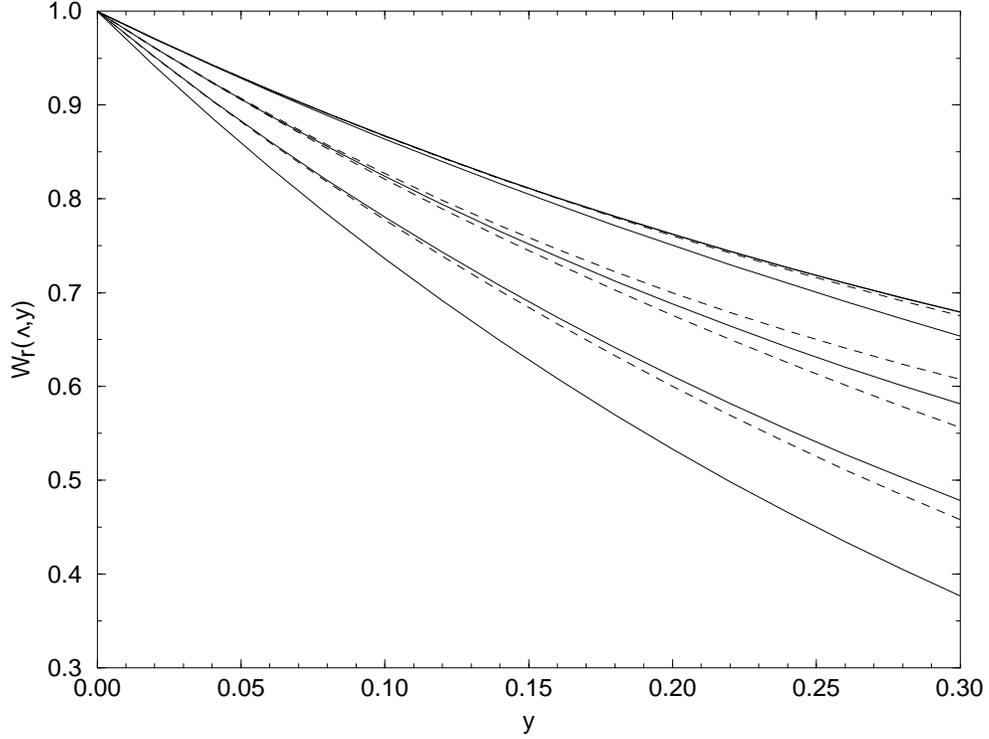}
\caption{Plots of $W_r(\Lambda,y)=q^{-1}W(\Lambda,y)$ as a function of
$y=1/(q-1)$ for Archimedean lattices $\Lambda$. 
For visual clarity, the curves are shown
alternately as solid (s) and dashed (d).  The order of the curves, from bottom
to top, is: $(3^6)$ (s), $(3^4 \cdot 6)$ (d), $(3^3 \cdot 4^2)$ (s), 
$(3 \cdot 6 \cdot 3 \cdot 6)$ (d), $(3 \cdot 4 \cdot 6 \cdot 4)$ (s), 
$(4^4)$ (d), $(3 \cdot 12^2)$ (s), $(6^3)$ (d), and 
$(4 \cdot 8^2)$ and $(4 \cdot 6 \cdot 12)$ as a single solid curve to this
resolution.  See text for further details.} 
\label{wrfig}
\end{figure}

\begin{figure}
\centering
\leavevmode
\epsfxsize=4.0in
\epsffile{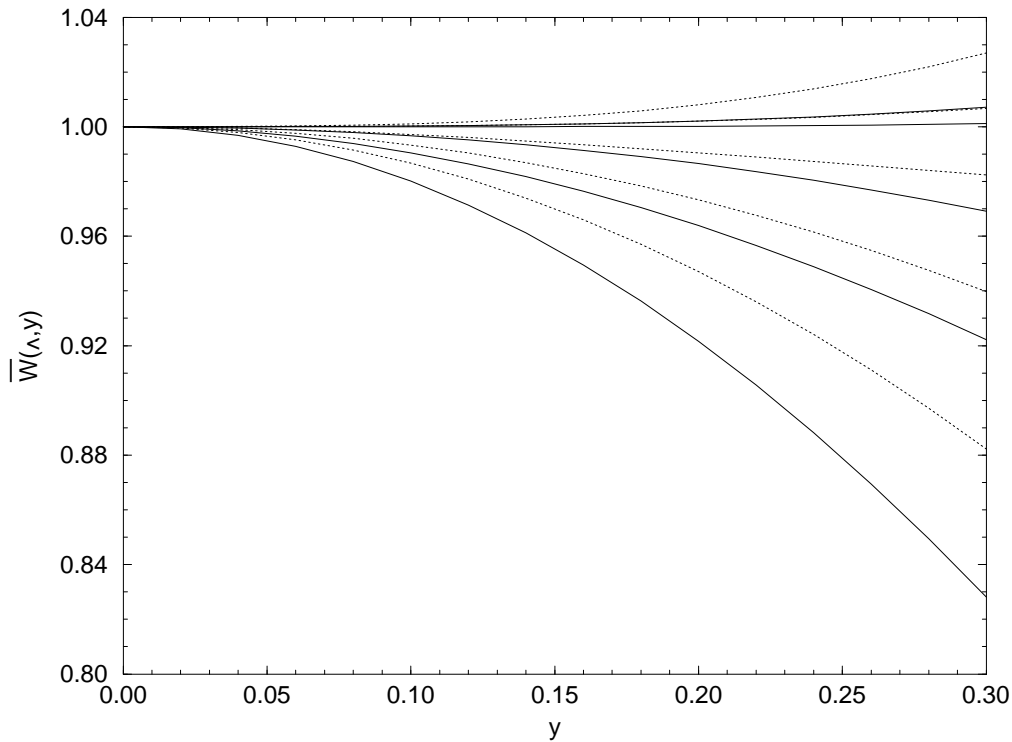}
\caption{Plots of $\overline W(\Lambda,y)$ as a function of $y=1/(q-1)$ for
Archimedean lattices $\Lambda$.  For visual clarity, the curves are shown
alternately as solid (s) and dotted (d).  The order of the curves, from bottom
to top, is: $(3^6)$ (s), $(3^4 \cdot 6)$ (d), $(3^3 \cdot 4^2)$ (s), 
$(3 \cdot 6 \cdot 3 \cdot 6)$ (d), $(3 \cdot 12^2)$ (s), 
$(3 \cdot 4 \cdot 6 \cdot 4)$ (d), $(6^3)$ (s), $(4 \cdot 8^2)$ and 
$(4 \cdot 6 \cdot 12)$ as almost coincident dotted and solid curves, and
finally, $(4^4)$ (d).  See text for further details.} 
\label{wbarfig}
\end{figure}

In Fig. \ref{wrfig} we plot $W_r(\Lambda,q)=q^{-1}W(\Lambda,q)$ for the eleven
Archimedean lattices, for $y=1/(q-1)$ in the range $0 \le y \le 0.30$,
corresponding to $q$ greater than about 4.  We have evaluated these from our
small-$y$ series expansions of $\overline W(\Lambda,y)$ together with the
definition (\ref{wbar}).  We find that $W_r(\Lambda,q)$ and $W(\Lambda,q)$, and
hence the ground state entropy of the $q$-state Potts antiferromagnet, 
$S_0(\Lambda,q)=k_B\ln W(\Lambda,q)$, are monotonically decreasing 
functions of the coordination number $\Delta(\Lambda)$ of the lattice
$\Lambda$.  This is a consequence of the fact that as one increases $\Delta$, 
one increases the number of constraints restricting the coloring of each 
vertex of the lattice.  Indeed, one can observe, especially for small $y$, that
the eleven curves in Fig. \ref{wrfig} fall into four groups, for the four
values of $\Delta=3,4,5$ and 6.  For further analytical purposes, we include
also a plot of the reduced function $\overline W(\Lambda,y)$ in
Fig. \ref{wbarfig}.  Again, this is calculated from our small-$y$ series.  From
our detailed comparison of lower bounds and small-$y$ series for the 
homopolygonal lattices and the $(4 \cdot 8^2)$ lattice with Monte Carlo
calculations, we expect that these curves are accurate over the range shown,
i.e., $y=0$ to $y=0.3$.  We have also checked this by evaluating the sizes of 
the highest calculated terms in our series.

   We note the following theorem:

\begin{flushleft} Theorem: If a lattice $\Lambda'$ can be obtained
 from another, $\Lambda$, by connecting disjoint vertices of $\Lambda$ with
bonds, then $W(\Lambda',q) \le W(\Lambda,q)$ (where $q \ge 0$ is an
integer). For $q \le \max\{q_c(\Lambda), q_c(\Lambda')\}$, this inequality
applies to the $W$ functions defined with the $D_{nq}$ order of limits in
eq. (\ref{wdefnq}), while for larger $q$, it applies to either of the 
definitions (\ref{wdefnq}) and (\ref{wdefqn}) since they are equivalent for 
this latter range of $q$. 

\end{flushleft}

\vspace{4mm}

\begin{flushleft}  Proof.  Consider a finite, $n$-vertex lattice of type 
$\Lambda$, denoted $\Lambda_n$.  We use the addition-contraction theorem from
graph theory, the statement of which is the following: let $G$ be a graph, and
consider any two vertices $v$, $v'$ on $G$ which are not adjacent, i.e., not 
connected by a bond of $G$.  Denote the graph with a bond connecting $v$ and
$v'$ as $G_{v-b-v'}$ and the graph with these two vertices identified as
$G_{v=v'}$.  Then $P(G,q) = P(G_{v-b-v'},q)+P(G_{v=v'},q)$.  Since 
$P(G_{v=v'},q) \ge 0$, it follows that $P(G_{v-b-v'},q) \le P(G,q)$. 
Now take $\Lambda_n = G$ and add bonds as necessary to construct the lattice 
$\Lambda'_n$.  Each time one adds a bond, one gets an inequality on the
correponding chromatic polynomials, thereby producing a sequence of such
inequalities, expressing the fact that the chromatic polynomial for the
original lattice is $\ge$ that for the lattice with one bond added, which is
$\ge $ that for the lattice with two bonds added, etc.  
Together, these yield the
inequality $P(\Lambda_n,q) \ge P(\Lambda'_n,q)$.  Now let $n \to \infty$ to
obtain $W(\Lambda,q)_{D_{nq}} \ge W(\Lambda',q)_{D_{nq}}$.  For 
$q > \max\{q_c(\Lambda), q_c(\Lambda')\}$, both of these definitions are
equivalent, so one can drop the subscripts $D_{nq}$ and $D_{qn}$. 

\end{flushleft}

   We give two examples of this theorem.  (Subscripts indicating orders of
limits in the definition of $W$ are understood where necessary.) 
First, as recalled above, the square lattice can be obtained from the 
honeycomb lattice by such bond addition, and consequently, 
$W((4^4),q) \le W((6^3),q)$.  Second, the triangular lattice can be obtained
 from the square lattice by bond addition, so $W((3^6),q) \le W((4^4),q)$.

   Concerning the plot of $\overline W(\Lambda,y)$, we observe one basic 
feature: as $y$ increases from 0, i.e. $q$ decreases from $\infty$, the initial
behavior of $\overline W(\Lambda,y)$ can be understood as resulting from the
leading term in the small-$y$ expansion 
\beq
\overline W(\Lambda,y) = 1 + (-1)^g \nu_g y^{g-1} + ... 
\label{wleading}
\eeq
where $g$ denotes the girth of $\Lambda$.  That is, for even (odd) $g$, 
$\overline W(\Lambda,y)$ increases (decreases).  

\vspace{4mm} 

   It is of interest to compare our results for the dependence of the
$W(\Lambda,q)$ coloring function, or equivalently, the exponent of the ground
state entropy of the Potts antiferromagnet, on the 
lattice coordination number $\Delta$ with the $\Delta$-dependence of other
models that exhibit ground state entropy.  We first consider the (spin 1/2) 
Ising antiferromagnet (IAF) on the triangular and kagom\'e lattices.  The 
exact solutions in these cases yield \cite{wannier}
\beq
S_0(IAF,(3^6))/k_B = \frac{1}{2}\int_{0}^{2\pi}\frac{d^2\theta}{(2\pi)^2}
 \ln(3+2P) = \frac{2}{\pi}\int_0^{\pi/3}d\omega \ln(2\cos\omega) \simeq 0.3231
\label{iafs0tri}
\eeq
for the triangular lattice, where
\beq
P = \cos(\theta_1) + \cos(\theta_2) + \cos(\theta_1+\theta_2)
\label{pfunction}
\eeq
and, for the kagom\'e lattice \cite{suto}
\beq
S_0(IAF,(3 \cdot 6 \cdot 3 \cdot 6))/k_B = 
\frac{1}{6}\int_0^{2\pi}\frac{d^2\theta}{(2\pi)^2} \ln(21-4P) \simeq 0.5018 
\label{iafs0kag}
\eeq
Thus, although the ground state entropy is accompanied by frustration in these
cases, in contrast to the $q$-state Potts antiferromagnet for the range $q \ge
\chi(\Lambda)$ considered in the rest of this paper, it is again true that the
ground state entropy decreases as the coordination number $\Delta$ of the
lattice increases.  A similar dependence has been reported in the case of
the quantum Heisenberg antiferromagnet; for the cases where this model involves
frustration, recent studies indicate that it has a ground state with 
long range order (albeit nonmaximal) on the triangular lattice, but with 
nonzero ground state entropy and no long range order in the case of
the kagom\'e lattice \cite{hafkag}. 

   We also make a comparison with the $\Delta$ dependence of generalized ice 
models.  At normal pressures $p \sim 1$ atm, ice forms a wurzite crystal with
fixed coordination number $\Delta=4$, so, of course, one cannot vary $\Delta$ 
in a realistic ice model.  A very accurate estimate of the entropy of ice was 
obtained by Pauling \cite{lp}: $S_0(ice)_{P,molar}/R = \ln(3/2) = 0.4055$, 
where, as above, $R=N_{Avog.}k_B$, to be compared with the measured value of 
$S_0/R = 0.41 \pm 0.03$ \cite{ice}.  Although one cannot vary 
$\Delta$ for real ice itself at $p=1$ atm., one can consider an abstract 
statistical model constructed to have two-valued variables (say arrows) 
assigned to links subject to the constraint of local arrow conservation, 
inspired by the physical and chemical constraint of local electric neutrality 
in real ice.  Clearly, such models can only be defined on a lattice with even 
coordination number $\Delta$. 
One sees that this immediately constitutes a difference with spin
models, which can be defined on lattices with even or odd $\Delta$, as well as
lattices such as the dual Archimedean lattices on which different vertices
$v_i$ have different degrees $\Delta_i$.  We recall that there is a 
straightforward generalization of the Pauling estimate for the exponential of
the entropy of real ice to that for an abstractly defined ice-type model; this
consists of the product $2^E$ describing the unconstrained number of 
positions of the hydrogen ions on the bonds 
(where $E=(\Delta/2)V$ and $V$ denotes the number of vertices on the lattice),
multiplied by a reduction factor which is the fraction of
the number of configurations for each vertex (oxygen location) that satisfy 
local electric neutrality. For each vertex this fraction is ${\Delta \choose
\Delta/2}/2^\Delta$, so combining these two factors yields the generalized
Pauling estimate for the exponent of the entropy, per site, for ice-type 
models (denoted by a subscript $I$), 
\beq
W_I(\Lambda)_P = 2^{-\Delta/2}{ \Delta \choose \Delta/2 }
\label{wip}
\eeq
This is a monotonically increasing function of (even) $\Delta$. Since
(\ref{wip}) is known to be a rigorous lower bound on $W_I(\Lambda)$ \cite{od},
it follows that $W_I(\Lambda)$ is also a monotonically increasing function of
(even) $\Delta$.  This shows 
that different models that both exhibit nonzero ground state entropy, such 
as (i) the zero-temperature Potts antiferromagnets for 
$q \ge \chi(\Lambda)$ on various lattices $\Lambda$; the (frustrated) Ising 
antiferromagnet; and the (frustrated) quantum Heisenberg antiferromagnet on 
the triangular and kagom\'e lattices on the one hand, and (ii) ice-type 
models on the other hand, can have quite different dependences on lattice 
properties such as the coordination number $\Delta$. 

\section{Chromatic Zeros and Support for a Conjecture for Regular Lattices}

\begin{figure}
\centering
\leavevmode
\epsfxsize=5.5in
\epsffile{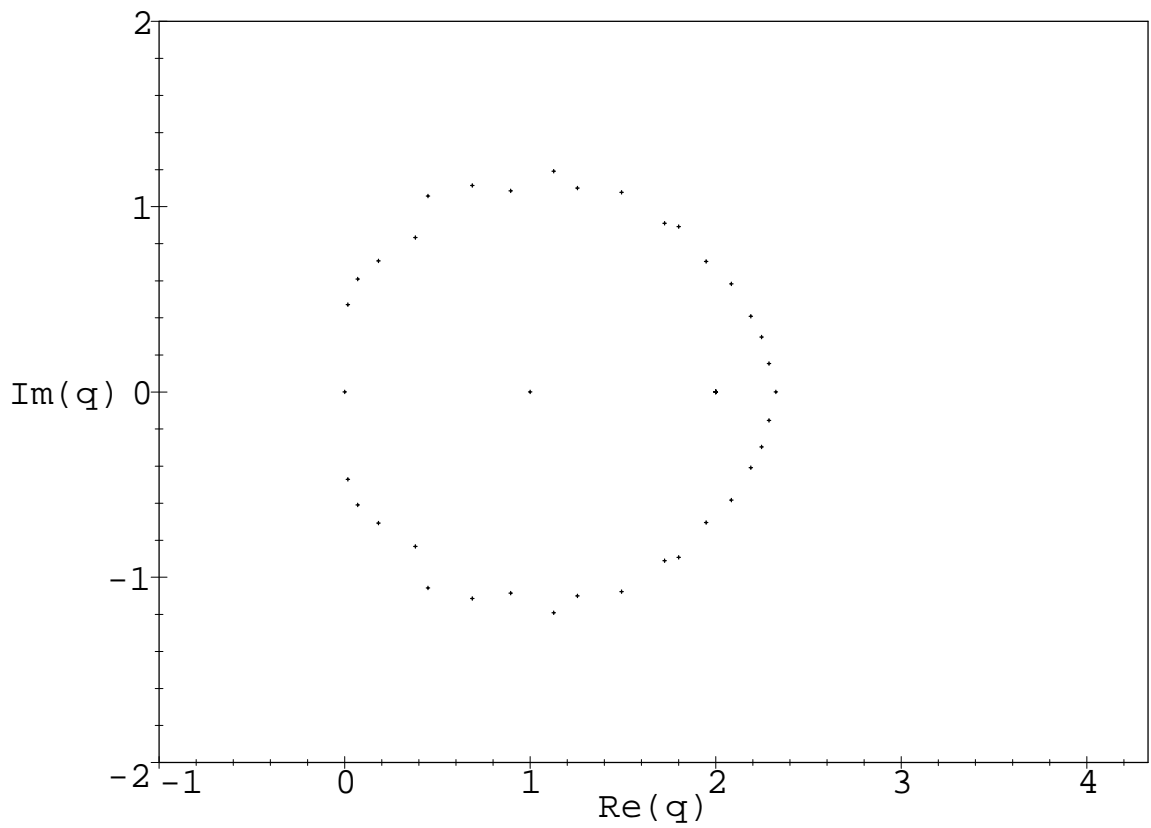}
\caption{Chromatic zeros for a section of the $(3 \cdot 12^2)$ lattice, with 
$n=48$ vertices.}
\label{cz31212fig}
\end{figure}

\begin{figure}
\centering
\leavevmode
\epsfxsize=5.5in
\epsffile{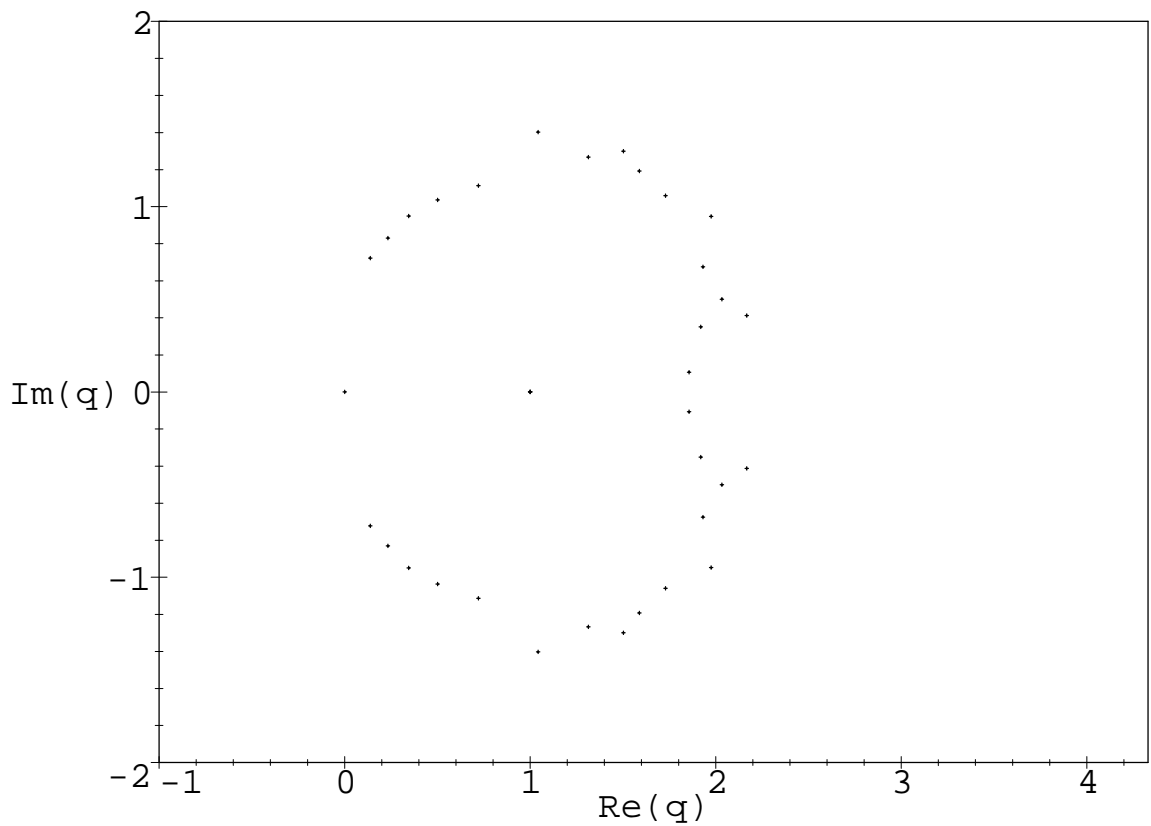}
\caption{Chromatic zeros for a section of the $(4 \cdot 8^2)$ lattice, with 
$n=36$ vertices.}
\label{cz488fig}
\end{figure}

   We have calculated chromatic polynomials for finite sections of 
heteropolygonal Archimedean lattices and for several Laves lattices, and in
each case have calculated the zeros of these polynomials, i.e., the respective
chromatic zeros for these lattices.  We have also done this for the simple
cubic lattice. 
Our main motivation for these calculations is to check whether the
results are consistent with the conjecture that we made previously \cite{wa}, 
namely that a sufficient condition that $W_r(\{G\},q)$ be analytic at $1/q=0$,
(i.e. that the region boundary ${\cal B}$ separating different regions where 
$W(\{G\},q)$ is analytic does not have any components that extend to complex 
infinity in the $q$ plane) is that $\{G\}$ is a regular lattice.  (This is 
not a necessary condition; as our exact results in Ref. \cite{w} showed, 
there are many families of graphs that are not regular lattices but do have 
compact region boundaries ${\cal B}$.)  In Refs. \cite{w,wa} we compared
the chromatic zeros for various families of graphs with exact results on the $n
\to \infty$ limits of these graphs and the corresponding $W(\{G\},q)$ 
functions, where $\{G\}$ denotes the $n \to \infty$ limit of $n$-vertex graphs
of type $G$.  As we discussed, as $n \to \infty$, the chromatic zeros (aside 
 from a well-understood discrete subset including zeros at $q=0,1$ and, for
graphs containing at least one triangle, also $q=2$) merge to form boundary 
curves ${\cal B}$ that separate regions of the complex $q$ plane in which
$W(\{G\},q)$ takes on different analytic forms.  For finite graphs, these
chromatic zeros lie near to, or, for some families of graphs \cite{w} 
exactly on, the asymptotic boundary curves ${\cal B}$.  However, for cases
where ${\cal B}$ does have components extending to complex infinity in the $q$
plane, we found that the chromatic zeros calculated on finite lattices 
deviate strongly from the parts of ${\cal B}$ that extend to infinity (see also
\cite{read91}).  In particular, as one increases the lattice size, one finds
that the (complex-conjugate) complex chromatic zeros farthest from the 
real axis in the $q$ plane move farther away from this axis.  
This, then, serves as a means by which one can infer, from 
calculations of chromatic zeros on finite lattices, whether the corresponding 
region boundaries ${\cal B}$ have components that extend to complex infinity in
cases where one does not have exact solutions for $W(\{G\},q)$ in 
$n \to \infty$ limit available.  As we have discussed before \cite{w,wa}, exact
results for the triangular lattice and chromatic zeros calculated on finite
square and honeycomb lattices \cite{baxter} yield boundary curves ${\cal B}$
that satisfy our conjecture.  Summarizing our results for various
Archimedean and dual Archimedean lattices, and for the simple cubic lattice, 
we find in all cases that the chromatic zeros are consistent with our 
conjecture in Ref. \cite{wa}.  
We show two typical examples in Figs. \ref{cz31212fig} and \ref{cz488fig}.  In
both cases, we use free boundary conditions and choose sections of the lattice
that have comparable lengths in the $x$ and $y$ directions.  

   From our earlier comparisons of chromatic zeros for various families of
graphs with the exact region boundaries ${\cal B}$ calculated in the limit of
infinitely many vertices \cite{w,wa}, we know that it is possible to make some 
reliable inferences about these boundaries from the positions of the chromatic
zeros for finite lattices.  Indeed, a subset of these chromatic zeros merge to
form the boundaries ${\cal B}$ in this limit.  (There are also discrete
isolated chromatic zeros such as those at $q=0,1$ and, if $\Lambda$ contains
triangles, also at $q=2$.)  Our calculations of chromatic zeros for
Archimedean lattices and their duals in 2D and for the simple cubic lattice are
consistent with the inference that in the thermodynamic limit, the respective
boundaries (1) separate the complex $q$ plane into at least two regions, one of
which (denoted $R_1$ in Refs. \cite{w,wa}) includes the positive real $q$ axis 
extending to $q=\infty$ and the circle at complex infinity, i.e., the image 
under inversion of the origin in the $1/q$ plane; and (2) the outermost 
component of the boundary ${\cal B}$ intersects the real $q$ axis on the left
at $q=0$ (for all $\Lambda$) and on the right at the point that we have 
denoted $q_c(\Lambda)$. Further studies on larger lattices will help to
elucidate the detailed shapes of the boundaries ${\cal B}$ for various
lattices.  For example, using sufficiently large lattices together with
comparisons of chromatic zeros for different lattices sizes to measure 
finite-size shifts of these zeros, one can carry out an extrapolation to the 
thermodynamic limit to determine the value of $q_c(\Lambda)$ with reasonable 
accuracy.  Work on this is in progress.

\section{Conclusions}

   The subject of nonzero ground state entropy is a fundamental one in
statistical mechanics.  In this paper we have proved a general rigorous lower
bound for $W(\Lambda,q)$, the exponent of the ground state entropy of the
$q$-state Potts antiferromagnet on an arbitrary Archimedean lattice.  The 
function $W(\Lambda,q)$ is also of considerable interest in mathematics, in
particular, the coloring of (finite) graphs and their infinite-$n$ limits. 
   From calculations of large-$q$ series expansions for the exact 
$\overline W(\Lambda,y)$ functions and comparison with our lower bounds on the 
various Archimedean lattices $\Lambda$, we have shown that the lower bounds are
actually very good approximations to the exact functions for large $q$. 
We have also calculated lower bounds and series for the duals of
Archimedean lattices.  Finally, from calculations of chromatic zeros on a
number of lattices, we have obtained further evidence for the conjecture that 
a sufficient condition for $q^{-1}W(\Lambda,q)$ to be analytic at $1/q=0$ is 
that $\Lambda$ is a regular lattice.

This research was supported in part by the NSF grant PHY-93-09888.

\vfill
\eject

\begin{thebibliography}{99}
\bibitem{lit}{Our results for $\chi(\Lambda)$ are not difficult to obtain,
although we have not been able to find them in the mathematical literature 
on graph theory, which deals mostly with proper (finite) graphs.}
\bibitem{comp}{The complete graph $K_r$ is defined as the graph consisting of
$r$ vertices each of which is connected to all of the others with bonds.}
\bibitem{wred}{The lower bound $W_r(\Lambda,q)_\ell \equiv 
q^{-1}W(\Lambda,q)_\ell$ for the reduced function $W_r(\Lambda,q)$ of eq. 
(\ref{wr}) is not, in general, monotonically increasing over this range 
$q \ge \chi(\Lambda)$, although 
$\lim_{q \to \infty} W_r(\Lambda,q)_\ell = 
\lim_{q \to \infty} W_r(\Lambda,q) = 1$.}
\bibitem{dk0}{In passing, we note also that from $P(C_{2k+1},q=2)=0$ and
eq. (\ref{dk}), it follows that $D_{2k+1}(q=2)=0$ for $k=1,2,...$.}
\end{thebibliography}
\end{document}